\documentclass[a4paper,10pt,twoside]{cpc-hepnp}

\usepackage{multicol}
\usepackage{multirow}
\usepackage{graphicx}
\usepackage{booktabs}
\usepackage{amssymb,bm,mathrsfs,bbm,amscd}
\usepackage[tbtags]{amsmath}
\usepackage{lastpage}

\begin{document}

\fancyhead[c]{\small Chinese Physics C~~~Vol. XX, No. X (201X)
XXXXXX} \fancyfoot[C]{\small 010201-\thepage}

\footnotetext[0]{Received 25 March 2015}

\title{Shape coexistence and evolution in neutron-deficient krypton isotopes\thanks{Supported by National Key Basic Research Program of China (2013CB83440) and National Natural Science Foundation of China (11235001 and 11320101004) }}

\author{
      BAI Zhi-Jun$^{1}$
\quad FU Xi-Ming$^{2}$
\quad JIAO Chang-Feng$^{3}$
\quad XU Fu-Rong$^{1;1)}$\email{frxu@pku.edu.cn}
}
\maketitle

\address{%
$^1$ State Key Laboratory of Nuclear Physics and Technology, School of Physics, Peking University, Beijing 100871, China\\
$^2$ National Institute for Radiological Protection, Chinese Center for Disease Control and Prevention, Beijing 100088, China\\
$^3$ Department of Physics and Astronomy, University of North Carolina, Chapel Hill, North Carolina, 27599-3255, USA\\
}

\begin{abstract}
Total Routhian Surface (TRS) calculations have been performed to investigate shape coexistence and evolution in neutron-deficient krypton isotopes ${}^{72,74,76}$Kr. The ground-state shape is found to change from oblate in ${}^{72}$Kr to prolate in ${}^{74,76}$Kr, in agreement with experimental data. Quadrupole deformations of the ground states and coexisting $0^{+}_{2}$ states as well as excitation energies of the latter are also well reproduced. While the general agreement between calculated moments of inertia and those deduced from observed spectra confirms the prolate nature of the low-lying yrast states of all three isotopes (except the ground state of ${}^{72}$Kr), the deviation at low spins suggests significant shape mixing. The role of triaxiality in describing shape coexistence and evolution in these nuclei is finally discussed.
\end{abstract}



\footnotetext[0]{\hspace*{-3mm}\raisebox{0.3ex}{$\scriptstyle\copyright$}2013
Chinese Physical Society and the Institute of High Energy Physics
of the Chinese Academy of Sciences and the Institute
of Modern Physics of the Chinese Academy of Sciences and IOP Publishing Ltd}%

\begin{multicols}{2}

\section{Introduction}
The equilibrium shape of an atomic nucleus is determined by a delicate interplay between macroscopic effects (including collective rotation) and microscopic effects (such as shell structure and Pauli blocking). One nucleus may have different shapes at different angular momenta and/or excitation energies. However, some nuclei may be found subject to a competition of distinct shapes that have the same spin and similar energies. This phenomenon, known as shape coexistence, has been of particular interest for a number of years~\cite{HW11}, since it is a sensitive probe to the nuclear quantum many-body correlations and serves as a testing ground for nuclear theories.

A variety of nuclear shapes are expected in the neutron-deficient $A\sim70$ mass region, making it an interesting mass region for the investigation of shape coexistence. The diversity is generally attributed to the abundance of low nucleon level densities, or ``shell gaps'' in this region. In fact, the relevant Nilsson diagram shows pronounced subshell gaps at nucleon numbers $34$ and $36$ (oblate), $34$ and $38$ (prolate), and $40$ (spherical). Therefore, adding or removing only a few nucleons might have a dramatic effect on nuclear shape. Moreover, in some cases, the competition of prolate, oblate, and spherical shapes is expected in one single nucleus.

Experimentally, the question of shape coexistence in neutron-deficient krypton isotopes arose more than three decades ago when irregularity was observed in the low-lying spectra of ${}^{74,76}$Kr and was interpreted in terms of a two-band mixing model~\cite{PHS81,PRH82}. Later, lifetime measurements of the same nuclei~\cite{GCC05} revealed that for both nuclei, the $2^{+}_{1}\rightarrow0^{+}_{1}$ transition has a significantly reduced $B(E2)$ value. This contrasts to the large collectivity manifested in other transitions of the yrast cascades, indicating considerable shape mixing at lower spins. Recently, a similar case was also reported on the neighboring nucleus ${}^{72}$Kr~\cite{ILM14}, where the $2^{+}_{1}\rightarrow0^{+}_{1}$ transition was found to be even more suppressed.

For even-even nuclei, more conclusive evidence of shape coexistence lies in the identification of low-lying excited $0^{+}$ states which could be seen as the ``ground states" of other shapes~\cite{AHD00}, or even better, rotational bands built on these excited $0^{+}$ states. For krypton, a candidate for such bands was early proposed in ${}^{76}$Kr~\cite{PHS81,PRH82}. In addition, a highly-deformed isomeric $0^{+}$ state in ${}^{74}$Kr was first hypothesized to explain the substantially ``prolonged" lifetime of the yrast $2^{+}$ state~\cite{CRP97}, and was soon corroborated by means of combined conversion-electron and $\gamma$-ray spectroscopy~\cite{BKH99}. Taking advantage of the same experimental technique, a low-lying isomeric $0^{+}$ state was later also established in ${}^{72}$Kr~\cite{BMK03}, extending shape coexistence to the $N=Z$ line.

As for what kinds of shape are involved in neutron-deficient krypton isotopes and how they evolve from one to another, some experimental efforts have been enlightening~\cite{CGK07}, while others failed to determine unambiguously whether the nuclei in question are prolate or oblate, even though they are known to be highly deformed~\cite{GBB05}. Alternatively, various theories have been applied to elucidate the detailed scenario, such as that employing Bohr's collective Hamiltonian~\cite{PFK83}, self-consistent triaxial mean-field models~\cite{BFH85,YMM01}, shell-model-based approaches~\cite{PSF96,PSF00,LDN03}, beyond (relativistic) mean-field studies~\cite{BBH06,FMX13}, and constrained Hartree-Fock-Bogoliubov (plus local Random-Phase-Approximation) calculations~\cite{GDG09,SH11}. In these calculations, the picture of shape coexistence can generally be reproduced, but more studies are needed to pin down the specifics, such as in which nuclei the transition of ground-state shape occurs, how large the deformations are, whether triaxiality plays a role, etc. The present work is aimed at shedding more light on these questions, by performing TRS calculations and scrutinizing both the ground-state properties and the rotational behaviors of these krypton isotopes.
\section{The Model}
The total-Routhian-surface (TRS) approach in the present work is a pairing-deformation-frequency self-consistent method based on the cranked shell model. In the TRS method~\cite{NWJ89,SWM94}, the total energy of a state consists of a macroscopic part that is given by the standard liquid-drop model~\cite{MS66}, a microscopic part that is calculated by the Strutinsky shell-correction approach~\cite{St67}, and the contribution due to rotation.

The single-particle levels which are needed in the calculation of the microscopic energy are obtained from the non-axially deformed Woods-Saxon potential~\cite{NDB85,CDN87}. Both monopole and quadrupole pairings are included. The monopole pairing strength $G$ is determined by the average-gap method~\cite{MN92}, and the quadrupole pairing strengths are obtained by restoring the Galilean invariance broken by the seniority pairing force~\cite{SK90,XSW00}. The quadrupole pairing has negligible effect on energy, but it is important for the proper description of moment of inertia (MOI)~\cite{SW95}. To avoid spurious pairing collapse at high angular momentum, an approximate particle-number projection is carried out by means of the Lipkin-Nogami method~\cite{PNL73}. In this method an extra Lagrange multiplier is introduced besides that in the BCS theory~\cite{BCS57}, in order to suppress particle number fluctuation.

The TRS calculations are performed in the three-dimensional deformation space $(\beta_{2},\gamma,\beta_{4})$. For a given deformation and rotational frequency, the pairings are self-consistently calculated by the HFB-like cranked-Lipkin-Nogami equation~\cite{SWM94}, so the dependence of pairing correlations on deformation and rotational frequency is properly treated. At each frequency, the equilibrium deformation of a rotational state is determined by minimizing the TRS in the deformation space $(\beta_{2},\gamma,\beta_{4})$.

\end{multicols}
\begin{center}
\includegraphics[width=16cm]{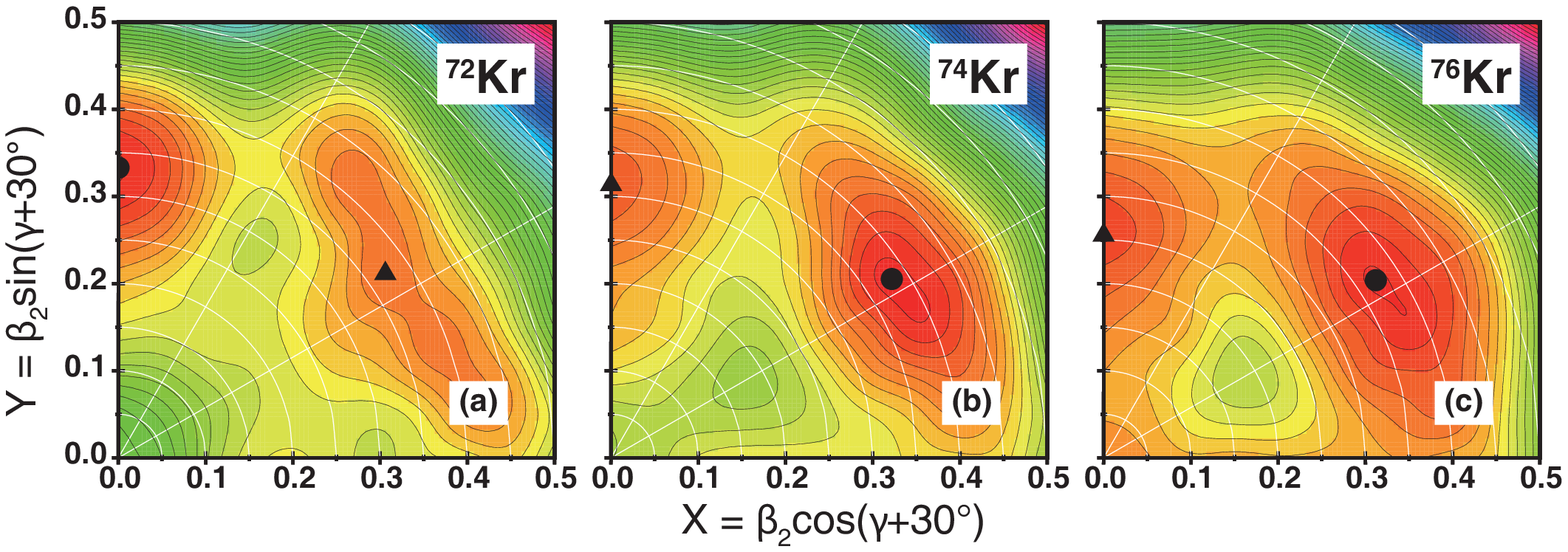}
\figcaption{\label{TRS0} (Color online) TRS's at $\omega=0$ for ${}^{72,74,76}$Kr. The solid circles and triangles denote the first and second lowest minima, respectively. The energy difference between neighboring contours is 200\,keV.}
\end{center}
\begin{center}
\tabcaption{ \label{tab}  Calculated deformation values of the ground states and the excited $0_{2}^{+}$ states of ${}^{72,74,76}$Kr, as well as energies of the latter, compared to available experimental data~\cite{GBB05,RNT01,NNDC}.}
\footnotesize
\begin{tabular*}{170mm}{@{\extracolsep{\fill}}cccccccccc}
\toprule Nuclide & \multicolumn{4}{c}{Ground state} & \multicolumn{3}{c}{$0^{+}_{2}$} & \multicolumn{2}{c}{$E(0^{+}_{2})$/MeV}\\\hline
 & $\beta_{2}$ & $\beta_{2}$(Exp.) & $\gamma$ & $\beta_{4}$ & $\beta_{2}$ & $\gamma$ & $\beta_{4}$ & TRS & Exp.~\cite{NNDC} \\ \cline{2-5}\cline{6-8}\cline{9-10}
${}^{72}$Kr & 0.333 & 0.330(21)~\cite{GBB05} & 60${}^{\circ}$ & 0.009 & 0.374 & \hphantom{6}5${}^{\circ}$ & \hphantom{$-$}0.026 & 0.673 & 0.6710(10)\hphantom{0} \\
${}^{74}$Kr & 0.381 & 0.419(25)~\cite{RNT01} & \hphantom{6}2${}^{\circ}$ & 0.013 & 0.318 & 60${}^{\circ}$ & \hphantom{$-$}0.002 & 0.547 & 0.509(1)\hphantom{000} \\
${}^{76}$Kr & 0.372 & 0.409(6)\hphantom{0}~\cite{RNT01} & \hphantom{6}3${}^{\circ}$ & 0.003 & 0.259 & 60${}^{\circ}$ & $-$0.013 & 0.307 & 0.76987(10) \\
\bottomrule
\end{tabular*}%
\end{center}
\begin{multicols}{2}
\section{Results and Discussion}
As is outlined in the \textit{Introduction}, for neutron-deficient krypton isotopes, experimental data concerning shape coexistence and evolution have been accumulating for the last three decades~\cite{ILM14,GCC05,PHS81,PRH82,CRP97,BKH99,BMK03,GBB05,CGK07}. In particular, a quadrupole deformation of $|\beta_{2}|=0.330(21)$ for the ground state of ${}^{72}$Kr has been deduced from the first determination of the absolute excitation strength $B(E2;0_{1}^{+}\rightarrow2^{+}_{1})$ in this nucleus~\cite{GBB05}. This value, though ambiguous by itself, yet in comparison with the predictions of a variety of self-consistent models, points to an oblate character of the ground state of ${}^{72}$Kr~\cite{GBB05}. 
As for ${}^{74,76}$Kr, low-energy Coulomb excitation experiments not only have confirmed the prolate nature of their ground-state bands, but also lend great support to the association of an oblate configuration with the bands built on excited $0_{2}^{+}$ states in spite of the complicated mixing scenario~\cite{CGK07}.

Fig.~\ref{TRS0} shows the TRS results for ${}^{72,74,76}$Kr (without cranking). Deformation values associated with the minima and excitation energies of the $0^{+}_{2}$ states are tabulated and compared to available experimental data in Table~\ref{tab}. Whereas all three isotopes exhibit distinct minima at both prolate ($\gamma\approx0^{\circ}$) and oblate ($\gamma\approx \pm 60^{\circ}$) deformations, the ground state, namely the lowest minimum shifts from oblate in ${}^{72}$Kr to prolate in ${}^{74,76}$Kr. This transition agrees perfectly with what various experiments have indicated, especially the Coulomb excitation experiments~\cite{GBB05,CGK07}. Besides, despite the fact that shape mixing is not taken into account in the present work, the calculated $E(0^{+}_{2})$ values are quite close to those given by measurement except for ${}^{76}$Kr. As for deformation, calculated $\beta_{2}$ for the ground state of ${}^{72}$Kr is in excellent agreement with the experimental value, and those of ${}^{74,76}$Kr are reasonably close. Moreover, it is worth noting that for the ground state of ${}^{74}$Kr $\beta_{2}\sim0.38$ is also one of the two experimental values~\cite{TCH90} before evaluation~\cite{RNT01}. Adopting this value and $\beta_{2}\sim0.32$ for the $0^{+}_{2}$ state of  ${}^{74}$Kr, the observed monopole strength $\rho^{2}(E0)$ could be well reproduced~\cite{CRP97}. This lends extra support to our calculations.

Thus, the present work has correctly located the nuclei where ground-state shape transition occurs and overall well reproduced excitation energies of the $0^{+}_{2}$ states and deformation values of the ground states. In contrast, beyond mean-field studies employing the Skyrme force~\cite{BBH06} predicted a dominance of oblate shape in the ground state all along the isotope chain, which contradicted the relevant experiments~\cite{CGK07}. The discrepancy was attributed partly to an intrinsic deficiency of the standard Skyrme interaction and partly to the limitation to axial symmetry.

The equilibrium shapes of nuclei could be attributed to a shell effect arising from low level densities around the Fermi levels of protons and/or neutrons. To illustrate this effect in the neutron-deficient krypton isotopes, we plot the single-particle level scheme for the neutrons of $^{72}$Kr in Fig.~\ref{72Kr-Nil}. That for the protons is basically the same, except for being a few MeV higher in absolute energy due to Coulomb repulsion. A pronounced shell gap, which is formed in the splitting of the $g_{9/2}$ orbital, is present for an oblate deformation of $\beta_2\sim -0.4$. This large gap gives rise to the oblate configurations of ${}^{72,74,76}$Kr (see Fig.~\ref{TRS0}). Likewise, the coexisting prolate configurations can also be associated with the corresponding shell gaps on the prolate side of the level scheme.

\begin{center}
\includegraphics[width=6cm]{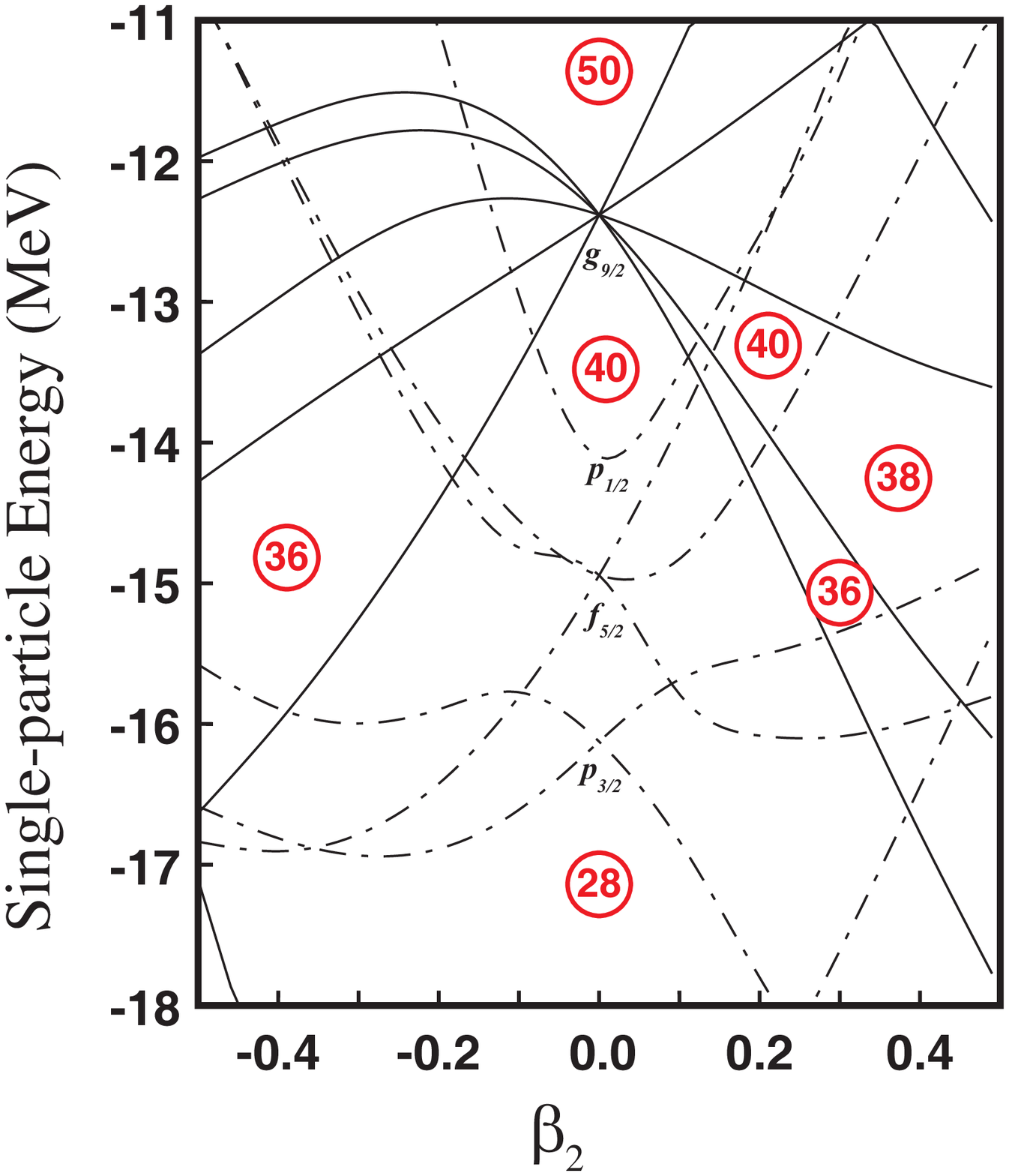}
\figcaption{\label{72Kr-Nil} (Color online) Nilsson diagram for ${}^{72}$Kr (neutrons), generated with the Woods-Saxon potential. Dashed lines denote levels with negative parity and solid ones those with positive parity. Diagrams for protons and for ${}^{74,76}$Kr are similar.}
\end{center}

In order to examine the deformation stability of these nuclei against rotation and investigate the behaviors of
rotational bands in the presence of shape coexistence, cranking calculations up to frequency $\hbar\omega\sim0.5$\,MeV have also been carried out. As an example, Fig.~\ref{74Kr-cranked} displays the obtained TRS for ${}^{74}$Kr that corresponds to a cranking frequency of $\hbar\omega=0.45$\,MeV. Compared to the TRS at $\hbar\omega=0$ (see panel (b) of Fig.~\ref{TRS0}), the minima stay much unchanged, except for being somewhat ``steeper", which is common, when a nucleus is cranked, due to the Coriolis force. This suggests that two low-lying rotational bands might be developed in ${}^{74}$Kr.

Fig.~\ref{J1-cmp} presents the calculation results for the kinematical MOI's corresponding to the predicted low-lying rotational bands of ${}^{72,74,76}$Kr, in comparison with available data~\cite{NNDC}. Apparently, our calculations support the prevalent view that ${}^{74,76}$Kr would make good prolate rotors except for the lowest levels. For both nuclei the experimental values of $\mathcal{J}^{(1)}$ start out small but then, with increasing angular momentum, converge to the gentle slopes of the prolate bands predicted by our calculations. The deviation at low frequencies can be qualitatively explained in terms of shape mixing. Combining Figs.~\ref{TRS0} and \ref{J1-cmp}, the ``pure" oblate band is generally higher in energy than its pure prolate competitor. Moreover, the difference in energy between competing levels of the same spin, which mix with each other in the actual spectrum, is bigger when spin is higher, leading to dampened mixing at higher spins. Therefore, while mixing always lowers the states of the prolate band in energy, the lowering is more significant at lower spins. The angular frequencies (kinematical MOI's) deduced from such a mixed spectrum are overestimated (underestimated) at low spins but become ``normal" when spin is high enough.

\begin{center}
  \includegraphics[width=6cm]{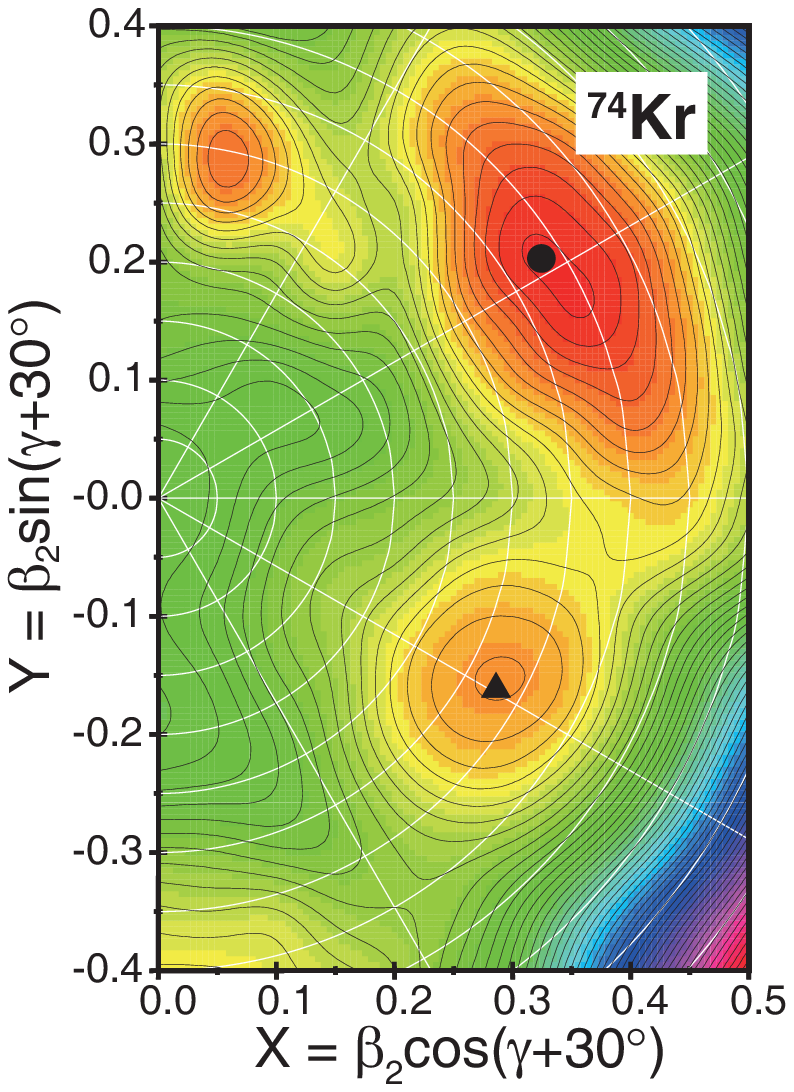}
  \figcaption{\label{74Kr-cranked} (Color online) Similar to Fig.~\ref{TRS0}, but for ${}^{74}$Kr at $\hbar\omega=0.45$\,MeV.}
\end{center}

What's more remarkable about Fig.~\ref{J1-cmp} is the ${}^{72}$Kr panel. The kinematical MOI comes quickly to coincide with the calculated prolate band, as in ${}^{74,76}$Kr. This confirms the prolate character of the upper members of the yrast line~\cite{AFG97,KWW01,PSF02,FLB03} but casts doubt on the designation of the ground state as the band head~\cite{AS10}. In fact, both experimental evidences~\cite{ILM14,GBB05} and theoretical calculations (ours as well as others'~\cite{GDG09,MSB09}) indicate that the ground state of ${}^{72}$Kr is more of an oblate nature. Besides, if the ground state is one member of the prolate band at all, since $E_{\gamma}(2_{1}^{+}\rightarrow 0^{+}_{1})=709.72(14)$\,keV is greater than $E_{\gamma}(4_{1}^{+}\rightarrow 2^{+}_{1})=611.68(14)$\,keV \cite{AS10}, the deduced angular frequency would decline a little before increasing steadily with angular momentum, which does not often occur in well-deformed nuclei. Some authors are inclined to take the excited $0_{2}^{+}$ state instead as the band head of the prolate band~\cite{BMK03}, which seems to make more sense in light of our calculation.

It has been indicated that the $\gamma$ degree of freedom might be of some importance in reproducing theoretically the observed phenomenon of shape coexistence and evolution in neutron-deficient krypton isotopes~\cite{BBH06,FMX13,GDG09}. In particular, beyond-mean-field studies adopting the Gogny D1S force~\cite{GDG09} can give a correct description only if triaxiality is included; otherwise, a dominance of oblate shape in the ground states would occur, similar to the results obtained by the studies that make use of the Skyrme interaction but are limited to axial symmetry~\cite{BBH06}. However, within the framework of the present work, although the equilibrium deformations seem to be soft in the $\gamma$ direction, especially for ${}^{72}$Kr (see Fig.~\ref{TRS0}), the calculation limited to axially-symmetric shapes would yield basically the same quadrupole deformations and excitation energies. Actually, for both non-cranking and cranking TRS's, the deviations from axial symmetry of the equilibrium shapes are nearly negligible (see Table~\ref{tab}). Nevertheless, some dynamic effects might be expected due to the $\gamma$ softness~\cite{FMX13}.

\end{multicols}

\begin{center}
\includegraphics[width=16cm]{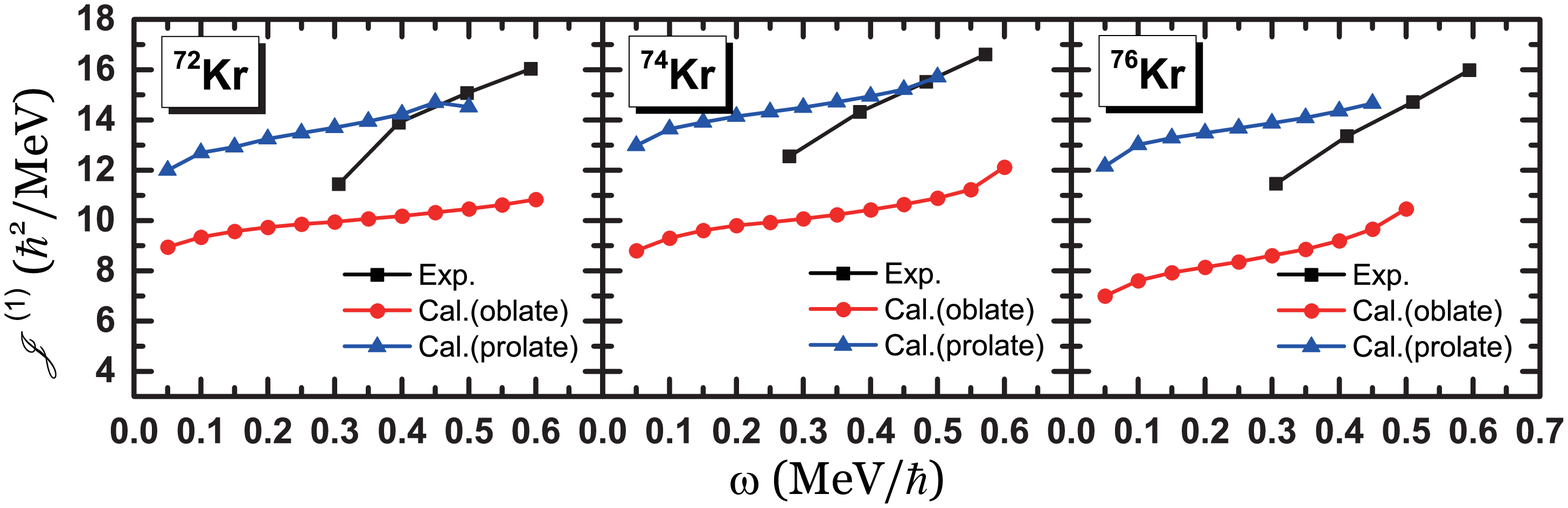}
\figcaption{\label{J1-cmp} (Color online) Comparison between calculated kinematical MOI's corresponding to predicted rotational bands of ${}^{72,74,76}$Kr and experimental values for ground-state bands~\cite{NNDC}.}
\end{center}

\begin{multicols}{2}

\section{Summary}

The TRS calculations have been performed for ${}^{72,74,76}$Kr. The phenomenon of shape coexistence and evolution in these nuclei is well described. In particular, we have successfully reproduced the ground-state shape transition, i.e. the ground state of ${}^{72}$Kr being oblate and those of ${}^{74,76}$Kr being prolate. The deformations and excitation energies given by our calculations are generally in agreement with measurement. The comparison between calculated and experimental moments of inertia not only confirms the prolate character of low-lying yrast states of all three isotopes (except the ground state of ${}^{72}$Kr), but also supports a picture of significant shape mixing at low spins. Finally, it is pointed out that, in contrast to that in the beyond-mean-field studies, triaxiality is not crucial in our description of shape coexistence and evolution in these krypton isotopes.

\acknowledgments{Discussion with LIU H L, Xi'an Jiaotong University is gratefully acknowledged.}

\end{multicols}

\vspace{10mm}


%
%
%
%
%
%
%
%
%

\vspace{-1mm}
\centerline{\rule{80mm}{0.1pt}}
\vspace{2mm}

\begin{multicols}{2}

\end{multicols}

\clearpage

\end{document}